\documentclass[a4paper]{article}
\usepackage[pdftex]{graphicx} 
\usepackage[english]{babel}
\usepackage[bf,small,nooneline]{caption}
\usepackage{multirow}
\usepackage[round]{natbib}

\title{Estimating Ion Escape from Unmagnetized Planets} 
\begin{document}
\date{April 7, 2021}
\author{M.\ Holmstr{\"o}m\thanks{Swedish Institute of Space Physics, PO~Box~812, SE-98128~Kiruna, Sweden. (\texttt{matsh@irf.se})}}
\maketitle

\begin{abstract}
  We propose a new method to estimate ion escape from unmagnetized planets that combines observations and models.  Assuming that upstream solar wind conditions are known, a computer model of the interaction between the solar wind and the planet is executed for different ionospheric ion production rates.
This results in different amounts of mass loading of the solar wind. 
  Then we obtain the ion escape rate from the model run that best fit observations of the bow shock location.  
  As an example of the method we estimate the heavy ion escape from Mars on 2015-03-01 to be $2\cdot 10^{24}$ ions per second, using a hybrid plasma model and observations by MAVEN and Mars Express.
  This method enables studies of how escape depend on different parameters, and also escape rates during extreme solar wind conditions, applicable to studies of escape in the early solar system, and at exoplanets.  
\end{abstract}

\section{Introduction}
Ion escape to space is important for the evolution of planetary atmospheres.
Neutrals in the upper parts of the atmosphere can be ionized by,
e.g., UV photons, charge exchange, and electron impacts.
The newly created ion can then be energized by electric fields and
transported away by the stellar wind, overcoming gravity,
resulting in atmospheric loss.

For planets in our solar system, we can observe the present day escape
of planetary ions by directly observing the ion flux near a planet.
This is done using an ion detector on a spacecraft and gives us the flux
of ions along the trajectory of the spacecraft.
Since the flux of escaping ions is highly variable in time and location,
accurately estimating the escape of ions can require observations
over many years to get an average escape rate.
To investigate how the escape rate of ions depend on different
parameters, e.g., upstream solar wind conditions, is even more difficult
due to the large amounts of observations needed to get sufficient statistics. 

Another way to estimate ion escape rates is to use computer models
of the solar wind interaction with a planet. 
An advantage of models compared to observations is that we at every instance
get the full three-dimensional escape.  In addition, there are no limitations
in sensitivity, energy range or field of view as there is for observations. 
However, to accurately estimate ion escape requires that the model contains
all important physical processes, in sufficient detail.
It is questionable if this is the case at present. 

Here we propose an alternative method of estimating ion escape.
A method that uses both observations and computer models. 
Instead of directly observing the escaping ions, we use observations of
other plasma quantities near the planet.  Then we use a parameterized model
and find the set of model parameters that give the best fit between
the model and the observations.  The model escape rate for the best fit
parameters then gives us an estimate of the ion escape rate.

This approach allow us to use data sets traditionally not used for
ion escape estimates, e.g., magnetic field and electron observations.
We can also estimate the escape rate from a very small set of observations,
during one orbit of a space craft around a planet, or during one flyby of a
planet.

To illustrate this general method we estimate the escape rate of
ions from Mars during one bow shock crossing of the MAVEN
spacecraft~\citep{Jakosky15} using
magnetic field observations of the bow shock location, 
observations of the upstream solar wind conditions, and a hybrid plasma
model.  We also use Mars Express (MEX)~\citep{Barabash07} observations
of electrons to verify our findings. 

The location and shape of the Martian bow shock has been the topic of
many observational studies.
Where is the bow shock located, and what are the controlling parameters?  
Recently \cite{Hall16} found a seasonal dependence of the bow shock location, and later also a solar cycle dependence~\citep{Hall19}.  The seasonal dependence should be due to the changing distance from the Sun of Mars.  It is not straight forward to deduce if it is due to changing solar wind pressure or changing EUV insolation, since both scale in the same way with distance from the Sun.  Also, both the solar wind and the EUV insolation changes over a solar cycle.
Regarding the effect of crustal magnetic fields on the bow shock location, \cite{Gruesbeck18} find such a dependence, and a modeling study by~\cite{Fang17} finds that a large part of the variability in escape may be due to the crustal fields, reducing and enhancing escape, depending on the location.
Regarding the shape of the bow shock, \cite{Vignes02} found that it is
furthest away from the planet in the hemisphere in the direction of the
solar wind convective electric field. 

However, it has been noted that, apart from the upstream solar wind conditions,
the factor controlling the location of the shock is the amount of mass
loading of the solar wind by ionospheric ions. 
This was noted for Venus by~\cite{Alexander85},
and later for Mars by~\cite{Vignes02} and~\cite{Mazelle04}.
The EUV flux, atmospheric state, ionospheric chemistry, magnetic anomalies,
and similar factors will all affect the location of the bow shock,
but only indirectly through the amount of mass loading.  
Therefore, given upstream solar wind conditions, we should be able to use
the amount of mass loading as a free parameter when modeling the location
of the bow shock.  

\section{Method}
We now describe the hybrid plasma solver used, the adaptation for Mars, the parameters used; the observations of magnetic field, ions and electrons used, and finally the algorithm to estimate ionospheric ion escape. 

\subsection{Hybrid model}
In the hybrid approximation, ions are treated as particles, 
and electrons as a massless fluid. 
The trajectories of the ions are computed from the Lorentz force, 
given the electric and the magnetic fields. 
The electric field is 
\begin{equation}
  \mathbf{E} = \frac{\displaystyle 1}{\displaystyle \rho_I} 
               \left( -\mathbf{J}_I\times\mathbf{B} 
  + \mathbf{J} \times \mathbf{B} 
   - \nabla p_e \right) + \eta \mathbf{J}, 
\label{eq:Ehyp}
\end{equation}
where $\rho_I$ is the ion charge density, 
$\mathbf{J}_I$ is the ion current density, 
$p_e$ is the electron pressure, and 
$\eta$ is the resistivity. 
The current is computed from, 
$\mathbf{J}=\mu_0^{-1}\nabla\times\mathbf{B}$, 
where $\mu_0=4\pi\cdot10^{-7}$ is the magnetic constant. 

Then Faraday's law is used to advance the magnetic field in time, 
        \[
          \frac{\displaystyle\partial \mathbf{B}}{\displaystyle\partial t} 
            = -\nabla\times\mathbf{E}. \label{eq:F}
        \]

Further details on the hybrid model used here, the discretization, 
and the handling of vacuum regions  
can be found in~\cite{Holmstrom12}. 

\subsection{Mars model}
In the hybrid simulation domain, Mars is modeled as a resistive sphere,
of radius $R$, 
centered at the origin, where all ions that hits the obstacle are removed
from the simulation.
The ionosphere is represented by the production of a single specie of ions
according to an analytical Chapman ionospheric profile~\citep{Holmstrom15}. 

We note that the ion production is a free parameter in such a model.
This is in contrast to models that self consistently include
ionospheric chemistry and a neutral atmosphere~\citep{Brecht16}.
Usually this free parameter is seen as a limitation of the model, but here 
we use this as an advantage, to find a
best fit to observations.  This means that it is not important
what the exact processes are in the ionosphere that produce the ions,
or how they are transported to the top of the ionosphere. 

Regarding the composition of escaping ionospheric ions.
A study by~\cite{Carlsson06} based on Mars Express observations found
flux ratios of O$_2^+$/O$^+$ = 0.9 and CO$_2^+$/O$^+$ = 0.2.
Later~\cite{Rojas18} estimated that O$_2^+$/O$^+$ = 0.76.
Using MAVEN data, \cite{Inui19} found a ratio of O$_2^+$/O$^+$ = 1.2, and that
CO$_2^+$ contributed less than 10\% to the total heavy ion flux.
In summary, observations indicate that the flux of escaping O$^+$ and O$_2^+$
ions are of similar magnitude, while the CO$_2^+$ flux is not significant. 

Since the code used here only handles one ionospheric ion specie,
we make separate simulation runs using O$^+$ and O$_2^+$
to investigate the effects of composition on the ion escape rate estimates. 

\subsection{Model parameters}
The coordinate system used is MSO coordinates with the solar wind
flowing along the $-x$-axis, with density $n_{sw}$, speed $v_{sw}$ and temperature $T_{sw}$.
The upstream interplanetary magnetic field is ${\bf B}_{sw}$. 
The computational grid has cubic cells of size $\Delta x$ and
the time step is $\Delta t$.
The computational domain is $-11000 \leq x \leq 10000$, $-33600 \leq y,z \leq 33600$ km. 
On the upstream boundary, after each timestep, we insert solar wind protons so that the number of particles per cell there is constant.
In the $y$ and $z$ directions we have periodic boundary conditions.  
The produced ionospheric ions has a weight (how many real ions they represent)
that is chosen such that the weight is similar to that of the solar wind protons. 
The model parameters and their values are listed in Table~\ref{tab:par}.

\begin{table}[htp]
  \begin{center}
    \caption{The parameters used for all simulation runs. }
    \label{tab:par}
    \begin{tabular}{l|c|c|c} 
      Name & Symbol & value & unit\\
      \hline
      Inner boundary radius & $R$ & 3540 & km\\
      Solar wind number density & $n_{\mathrm{sw}}$ & 2.4 & cm$^{-3}$\\
      Solar wind velocity & $u_{\mathrm{sw}}$ & 350 & km/s\\
      Solar wind temperature & $T_{\mathrm{sw}}$ & $1.2\cdot 10^5$ & K\\
      Solar wind magnetic field & ${\bf B}_{sw}$ & $(-1,-2.7,-1)$ & nT\\
      Plasma resistivity & $\eta$ & $5\cdot 10^4$ & $\Omega$ m\\
      Obstacle resistivity & & $7\cdot 10^5$ & $\Omega$ m\\
      Particles per cell &  & 128 & \\
      Weight of ionospheric ions & & $2.2\cdot 10^{21}$ & \\
      Height of max production & & 500 & km\\
      Atmospheric scale height & & 250 & km\\
      Cell size          & $\Delta x$ & 350 & km \\
      Time step          & $\Delta t$ & 0.05 & s \\
    \end{tabular}
  \end{center}
\end{table}

\subsection{Observations}
We use MAVEN magnetic field~\citep{Connerney15} and ion~\citep{Halekas17}
observations to determine upstream conditions and bow shock location.
The orbit is chosen such that the solar wind conditions are steady,
so that the conditions should be unchanged while MAVEN is inside the bow
shock, when we do not have observations of the solar wind.
This also allow us to use a simulation that does not have time dependent
upstream solar wind conditions.  
To verify our results we also use Mars Express (MEX) observations of
electrons~\citep{Frahm06} to locate bow shock crossings. 
The upstream solar wind parameters are listed in Table~\ref{tab:par}.

\subsection{Algorithm}
\begin{figure}
  \begin{center}
    \noindent\includegraphics[width=22pc]{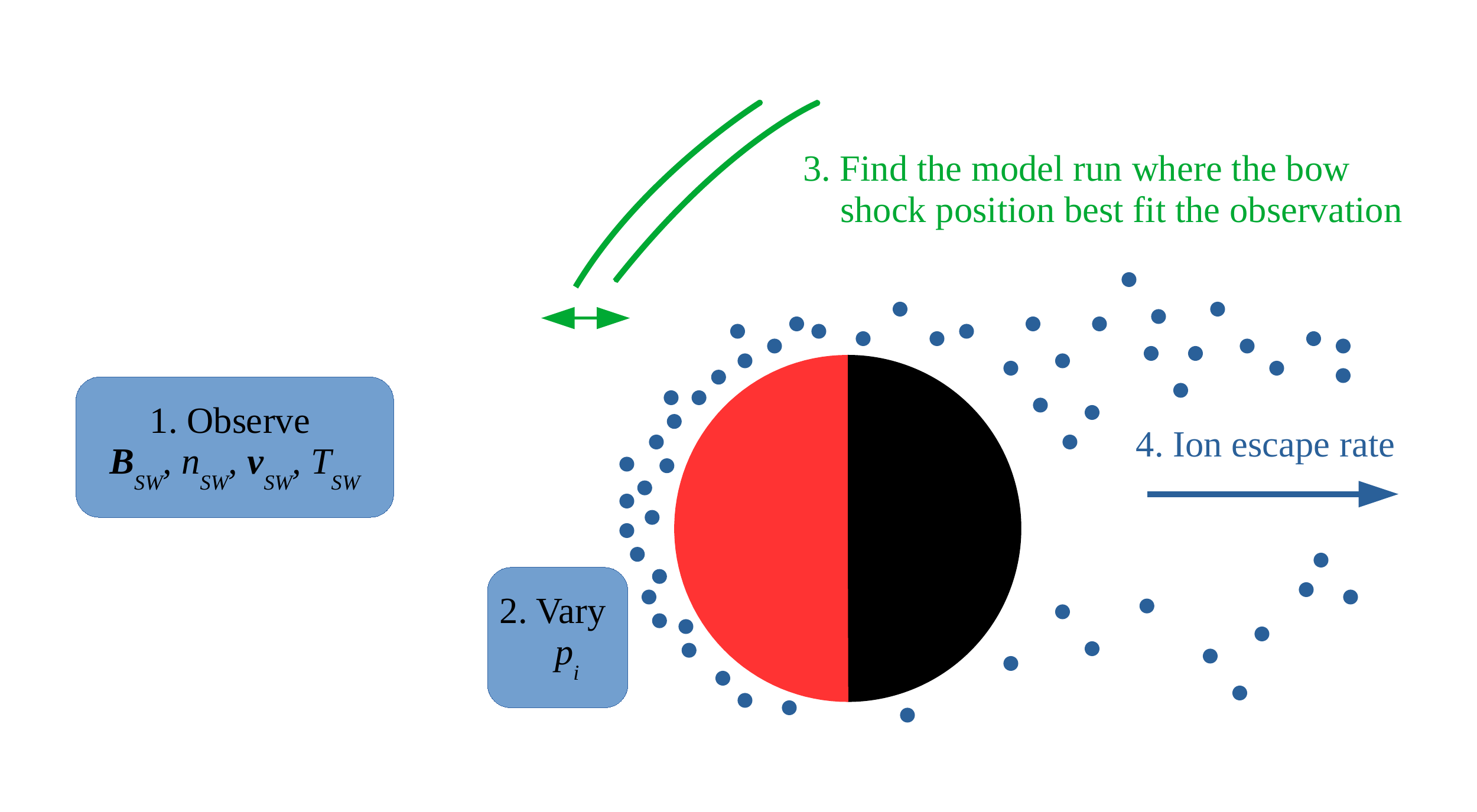}
    \caption{An illustration of the algorithm to estimate the ion escape rate.
      The Sun is to the left and Mars is the red and black disk. Using fixed upstream solar wind conditions from observations in the hybrid model, we vary the ionospheric heavy ion (blue dots) production rate for different simulation runs.  The bow shock location (green lines) in each simulation run is then compared to the observed bow shock location.  The estimated escape rate will then correspond to that of the simulation run that best fit the bow shock location. 
            } \label{fig:alg}   
  \end{center}
\end{figure}

Now we describe the algorithm for estimating the escape of ionospheric ions
from observations of the upstream solar wind and the location of the
bow shock.  The algorithm is illustrated in Fig.~\ref{fig:alg}.
\begin{enumerate}
\item We start with an observed state of the upstream solar wind: 
  The magnetic field; the solar wind density, velocity and temperature.
\item Then we make several runs of the hybrid model for these upstream solar wind conditions, using different ionospheric ion production rates.
\item We then find the simulation run that has a bow shock location that best correspond to the observed location.  This could be done quantitatively, e.g., by a least square fit, but here we visually compare the simulations and the observations.
  \item The escape for the best fit simulation run is then computed and this will be our estimate of the escape rate of ionospheric ions at the time of the bow shock observation. 
\end{enumerate}

\section{Results}
\begin{figure}
  \begin{center}
    \noindent\includegraphics[width=40pc]{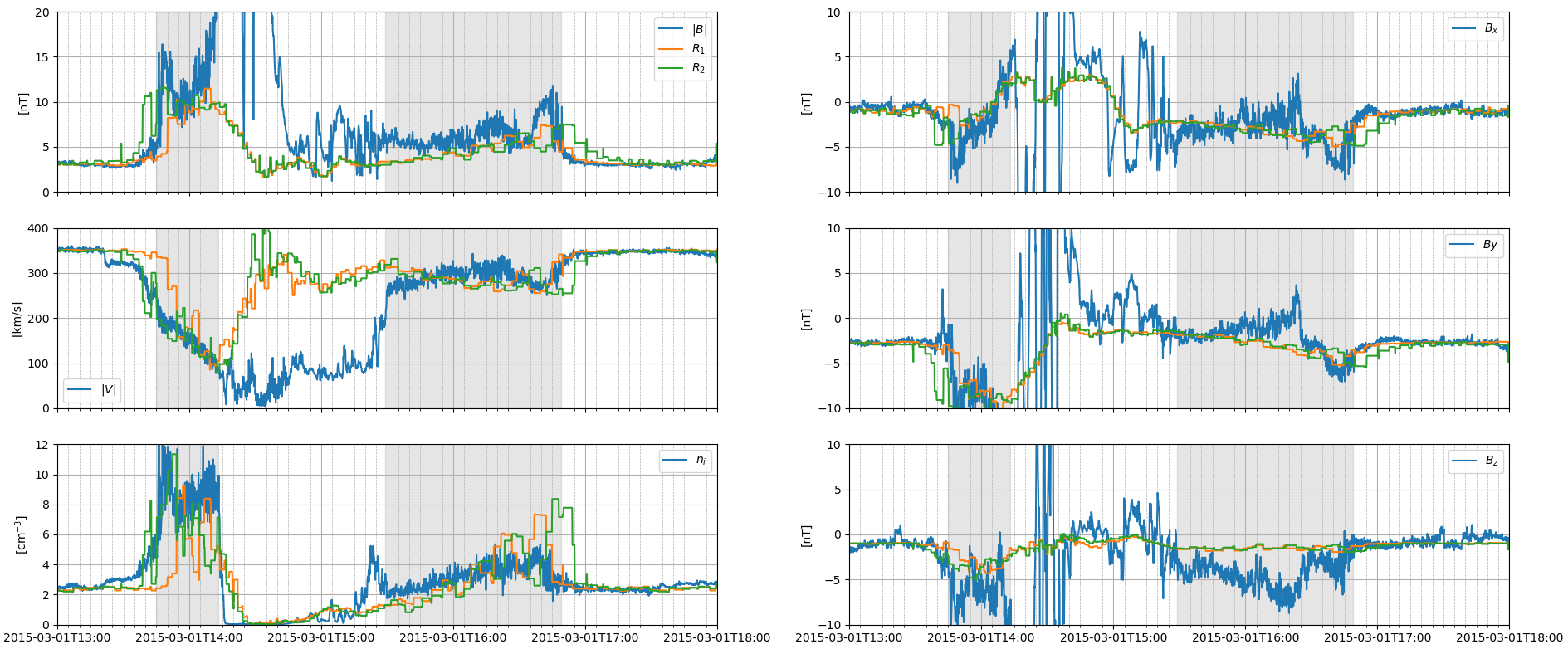}
    \caption{A comparison of MAVEN observations (blue) with
      two O$^+$ model runs ($R_1$ in orange and $R_2$ in green) at 490~s.  In the left column, from the top, we have magnetic field magnitude, proton velocity, and proton number density. In the right column we have the three magnetic field components ($B_x$, $B_y$, and $B_z$) in MSO coordinates.
      Indicated in gray is also the location of the magnetosheath as seen
      from the observations.  The induced magnetosphere is thus between the two gray regions. 
            } \label{fig:cmp}   
  \end{center}
\end{figure}

In Fig.~\ref{fig:cmp} we present a comparison of MAVEN observations and
two hybrid runs with an O$^+$ ionosphere that best fit the observations. 
We see that there is a fairly good agreement between the models and the
observations at the bow shock and in the magnetosheath.
Closer to the planet, in the induced magnetosphere, the agreement is
however not so good.  The magnetic field is much larger and more variable
than in the model.   This is not surprising since we have a model with
a very simplified ionosphere, no magnetic anomalies, and low spatial resolution.
Also, in the magnetosphere the proton velocities in the model are much higher than observed.  The proton density is however very small in this region.
Also, the variability of $B_z$ in the magnetosheath is smaller in the models
than observed. 

As expected, and seen when comparing the two model runs,
the location of the bow shock moves outward when the
ionospheric ion production is increased, resulting in a larger mass loading. 

The escape rate never reach a steady state due to intrinsic variability of the induced magnetosphere.  So we determine the escape rate by averaging the flux of ionospheric ions along $-x$ in the simulation domain downstream of $x=-5000$~km. This is done at 10~s intervals from 200~s to 590~s, and then we average the escape over those times.

In Table~\ref{tab:prod} we show the results for the different simulation runs,
numbered 1-4.
For the best fit runs using O$^+$ the escape rate is $2.0\cdot 10^{24}$,
while it is $1.5\cdot 10^{24}$ s$^{-1}$ for O$_2^+$.
Since we in reality has a mixture of the two ion species,
the escape rate should be between these values, and we can estimate
the actual escape as $2\cdot 10^{24}$ s$^{-1}$. 

\begin{table}[htp]
  \begin{center}
    \caption{The different best fit simulation runs,
      the ionospheric ion specie,
      the maximum ionospheric production rates, and
      the resulting escape rates. }
    \label{tab:prod}
    \begin{tabular}{l|c|c|c|c}
      Run & 1 & 2 & 3 & 4  \\
      \hline
      Ionospheric specie & O$^{+}$ & O$^{+}$ & O$_2^{+}$ & O$_2^{+}$ \\
      Production Rate, $p_i$ [cm$^{-3}$s$^{-1}$] & 0.4 & 0.5 & 0.4 & 0.5 \\
      Escape Rate [$10^{24}$s$^{-1}$] & 1.985 & 1.989 & 1.452 & 1.463 \\
    \end{tabular}
  \end{center}
\end{table}


We can note how the bow shock location depends on the specie of the escaping
ionospheric ions. A similar location is obtained with 25\% less escaping
O$_2^+$ ions compared to O$^+$.  It is not however directly proportional to the
total mass of the escaping ions, in that case we would expect a 50\% reduction.
So it is not only the amount of mass loading that determine the
bow shock location, the dynamics of the escaping ions is also important. 

Looking at the escape in Table~\ref{tab:prod} for the same spice, but for
different production rates, we see that the bow shock location is
very sensitive to the escape rate.  Less than 1\% variation in escape
results in the change in bow shock location clearly visible in
Fig.~\ref{fig:cmp}.  Another way to state this is that the
escape rate is weakly dependent on the production rate.  

\begin{figure}
  \begin{center}
    \noindent\includegraphics[width=30pc]{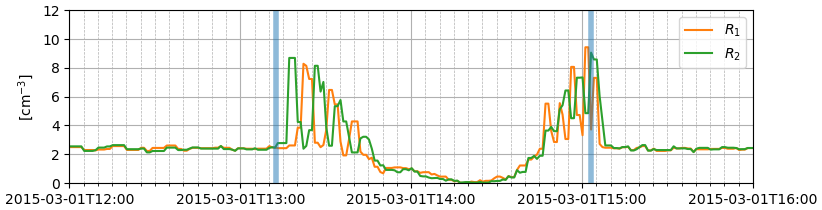}
    \caption{The proton number density along the MEX orbit for the two best fit simulation runs ($R_1$ in orange and $R_2$ in green), at 490 seconds of simulation time. The blue vertical lines show the two bow shock crossings seen in MEX electron data.  
            } \label{fig:cmp2}   
  \end{center}
\end{figure}

To verify the location of the bow shock in the two best fit model runs,
we also use Mars Express observations of the bow shock in electron data.
In Fig.~\ref{fig:cmp2} we plot the proton number density from the
same two hybrid runs as in Fig.~\ref{fig:cmp}, but now along the
Mars Express orbit, together with the location of the bow shock
crossings observed by MEX.
The agreement is fairly good, even if the observed bow shock is
a few minutes earlier than seen in the model runs.
This corresponds to a distance of a few hundred kilometer, comparable
to the simulation grid cell size.  
One reason for this could be that we have not used an aberrated solar
wind velocity (it flows along the $-x$ axis).
This should result in a tilt of the whole magnetosphere and bow shock.  

\section{Discussion}
For our example case we found an escape rate of $2\cdot 10^{24}$ ions per
second.  This is in the range of recently published estimates for
the escape rate at Mars~\citep{Ramstad15,Brain15,Dong17}.

Note that there is no neutral H or O corona in the model.  So those 
populations of exospheric pick-up ions are missing.  That means the
mass loading due to the corona is missing in the simplified model we
use.  The total mass loading can be compensated by more ions from
the ionospheric source, but in reality the spatial distribution
would be different.  We also do not include any alpha particles in the
solar wind, that should have some effect on the solar wind interaction. 

An assumption is that, given upstream condition, the mass loading determines the bow shock location.  This means that the proposed method is applicable to unmagnetized planets.  For magnetized planets the bow shock location is mainly determined by the upstream solar wind and the strength of the dipole field.  Escape at magnetized planets occur in the cusp regions, and how that affects the bow shock location, and if the presented method could be adapted to magnetized planets would need further investigation.  

The bow shock location has been found to depend on the location of the
magnetic anomalies relative to the solar wind flow~\citep{Fang17}.
Is it because the fields “push out” the boundaries, or because the fields
increase ion escape?  The latter may not require crustal fields in the model
used in our algorithm.  Since the parameter we vary is the amount of
ions near Mars that is available to escape.
If the crustal fields in a specific geometry enhance escape, this will
be captured in the algorithm since the best fit bow shock will be further
out, and vice versa.  

We used a hybrid plasma model in the algorithm.  It would however be possible to use another type of plasma model, e.g., a magnetohydrodynamic (MHD) model, that can predict the location of the bow shock for different amounts of mass loading by ionospheric ions, given upstream solar wind conditions.

\section{Conclusions} 
In the past, ion escape has been estimated either by computer models or from observations.  Models have the problem that every physical process has to be present in the model.  Observations suffer from variability, requiring the averaging of data over years, or even decades.  The method proposed here uses a model and observations together.  In that way we overcome the difficulties of each approach. We do not need to get every physical process correct in the model, and we can get an estimate of the escape using just one observation.  We use a model to estimate a global property (ion escape) from a local observation (bow shock location). Since we use a model, there are no physical limitations in terms of energy coverage and field of view that observations have.  In particular low energy escaping ions are difficult to observe.  

This opens up the possibility of estimating escape during flybys of unmagnetized planets, in the past and in the future.  It also allow for estimating escape twice per orbit given only a magnetometer and an ion detector.  This enables detailed studies of how the escape depends on different parameters.  Something that has been difficult in the past due to the years of observations needed to collect enough statistics. It also makes possible the study of escape during transient events, like extreme solar wind conditions, something that is important for studies of escape in the past, and the escape at exoplanets.

\section*{Acknowledgements}
Computing resources used in this work were provided by
the Swedish National Infrastructure for Computing (SNIC) at the
High Performance Computing Center North (HPC2N), Ume\aa\ University, Sweden.
The software used in this work was in part developed by the 
DOE NNSA-ASC OASCR Flash Center at the University of Chicago. 

The \mbox{ASPERA-3} electron data used, from the ELS sensor, is available in ESA's Planetary Science Archive (PSA) at \texttt{ftp://psa.esac.esa.int/pub/\\ mirror/MARS-EXPRESS/ASPERA-3/MEX-M-ASPERA3-3-RDR-ELS-EXT5-V1.0/}

The MAVEN data used in this work, ion data from the SWIA instrument and magnetic field data from the MAG instrument, is available in NASA's Planetary Data System (PDS) at \texttt{https://pds-ppi.igpp.ucla.edu/search/view/\\ ?f=null\&id=pds://PPI/maven.insitu.calibrated/data/2015/03}


\begin{thebibliography}{}

\bibitem[{Alexander and Russell}(1985)]{Alexander85}
  Alexander, C.J., and C.T Russell (1985),
  Solar Cycle Dependence of the Location of the Venus Bow Shock,
  \textit{Geophysical Research Letters},
  v. 12, n. 6, 369--371. https://doi.org/10.1029/GL012i006p00369

\bibitem[{Barabash et~al.}(2007)]{Barabash07}
  Barabash, S., R. Lundin, H. Andersson, K. Brinkfeldt, A. Grigoriev,
  H. Gunell, et al. (2007), The Analyzer of Space Plasmas and Energetic Atoms
  (ASPERA-3) for the Mars Express mission, \textit{Space Sci. Rev.},
  126(1--4), 113--164. https://doi.org/10.1007/s11214-006-9124-8

\bibitem[{Brain et~al.}(2015)]{Brain15}
  Brain, D. A., J.P. McFadden, J.S. Halekas, J.E.P. Connerney, S.W. Bougher,
  S. Curry, C.F. Dong et al. (2015), 
  The spatial distribution of planetary ion fluxes near Mars observed by MAVEN, 
  \textit{Geophysical Research Letters},
  42, 9142--9148. https://doi.org/10.1002/2015GL065293

\bibitem[{Brecht et~al.}(2016)]{Brecht16}
  Brecht, S.H., S.A. Ledvina, and S.W. Bougher (2016),
  Ionospheric loss from Mars as predicted by hybrid particle simulations,
  \textit{Journal of Geophysical Research: Space Physics},
  121, 10, 190--10. https://doi.org/10.1002/2016JA022548

\bibitem[{Carlsson et~al.}(2006)]{Carlsson06}
  Carlsson, E., A. Fedorov, S. Barabash, E. Budnik, A. Grigoriev,
  H. Gunell, et al. (2006),
  Mass composition of the escaping plasma at Mars, 
  \textit{Icarus}, 182(2):320--328. https://doi.org/10.1016/j.icarus.2005.09.020

\bibitem[{Connerney et~al.}(2015)]{Connerney15}
  Connerney, J.E.P., J.R. Espley, G.A. DiBraccio, J.R. Gruesbeck,
  R.J. Oliversen, D.L. Mitchell, et al. (2015),
  First results of the MAVEN magnetic field investigation,
  \textit{Geophys. Res. Lett.}, 42, 8819--8827. https://doi.org/10.1002/2015GL065366
  
\bibitem[{Dong et~al.}(2017)]{Dong17}
  Dong, Y., X. Fang, D.A. Brain, J.P. McFadden, J.S. Halekas,
  J.E.P. Connerney, et al. (2017),
  Seasonal variability of Martian ion escape through the plume and tail
  from MAVEN observations,
  \textit{Journal of Geophysical Research: Space Physics},
  122. doi:10.1002/2016JA023517

\bibitem[{Fang et~al.}(2017)]{Fang17}
  Fang, Xiaohua, Yingjuan Ma, Kei Masunaga, Yaxue Dong, David Brain,
  Jasper Halekas, et al. (2017),
  The Mars crustal magnetic field control of plasma boundary locations and
  atmospheric loss: MHD prediction and comparison with MAVEN, 
  \textit{Journal of Geophysical Research: Space Physics},
  122. doi:10.1002/2016JA023509.

\bibitem[{Frahm et~al.}(2006)]{Frahm06}
  Frahm, R.A., J.D. Winningham, J.R. Sharber, J.R. Scherrer, S.J. Jeffers,
  A.J. Coates, et al. (2006),
  Carbon dioxide photoelectron energy peaks at Mars,
  \textit{Icarus}, 182, 2, 371--382. https://doi.org/10.1016/j.icarus.2006.01.014
  
\bibitem[{Gruesbeck et~al.}(2018)]{Gruesbeck18}
  Gruesbeck, Jacob R., Jared R. Espley, John E.P. Connerney,
  Gina A. DiBraccio, Yasir I. Soobiah, David Brain, et al. (2018),
  The Three-Dimensional Bow Shock of Mars as Observed by MAVEN, 
  \textit{Journal of Geophysical Research: Space Physics},
  123. https://doi.org/10.1029/2018JA025366

\bibitem[{Halekas et~al.}(2017)]{Halekas17}
  Halekas, J.S., S. Ruhunusiri, Y. Harada, G. Collinson, D. L. Mitchell,
  C. Mazelle, et al. (2017), Structure, dynamics, and seasonal variability
  of the Mars-solar wind interaction: MAVEN solar wind ion Analyzer
  in-flight performance and science results,
  \textit{J. Geophys. Res. Space Physics},
  122, 547--578. https://doi.org/10.1002/2016JA023167
  
\bibitem[{Hall et~al.}(2016)]{Hall16}
  Hall, B.E.S., M. Lester, B. Sánchez-Cano, J.D. Nichols, D.J. Andrews,
  N.J.T. Edberg, et al. (2016), 
  Annual variations in the Martian bow shock location as observed
  by the Mars Express mission, 
  \textit{Journal of Geophysical Research: Space Physics},
  121, 11,474--11,494. doi:10.1002/2016JA023316
  
\bibitem[{Hall et~al.}(2019)]{Hall19}
  Hall, B.E.S., B. Sánchez‐Cano, J.A. Wild, M. Lester, and M. Holmström (2019), 
  The Martian Bow Shock Over Solar Cycle 23–24 as Observed by the Mars Express
  Mission, 
  \textit{Journal of Geophysical Research: Space Physics},
  124, 4761--4772. https://doi.org/10.1029/2018JA026404
  
\bibitem[{Holmstrom et~al.}(2012)]{Holmstrom12}
  Holmstr\"{o}m, M., S. Fatemi, Y. Futaana, and H. Nilsson (2012),
  The interaction between the Moon and the solar wind,
  \textit{Earth Planets Space}, 64(2):237–245. doi:10.5047/eps.2011.06.040

\bibitem[{Holmstrom and Wang}(2015)]{Holmstrom15}
  Holmstrom, Mats, and Xiao-Dong Wang (2015), 
  Mars as a comet: Solar wind interaction on a large scale,
  \textit{Planetary and Space Science}, 119, 43--47. https://doi.org/10.1016/j.pss.2015.09.017
  
\bibitem[{Inui et~al.}(2019)]{Inui19}
  Inui, S., K. Seki, S. Sakai, D.A. Brain, T. Hara, J.P. McFadden,
  et al. (2019),
  Statistical Study of Heavy Ion Outflows From Mars Observed in
  the Martian-Induced Magnetotail by MAVEN,
  \textit{Journal of Geophysical Research: Space Physics},
  Volume 124, Issue 7, pp. 5482--5497. https://doi.org/10.1029/2018JA026452

\bibitem[{Jakosky et~al.}(2015)]{Jakosky15}
  Jakosky, B. M., J. M. Grebowsky, J. G. Luhmann, and D. A. Brain (2015),
  Initial results from the MAVEN mission to Mars,
  \textit{Geophys. Res. Lett.}, 42, 8791--8802.  https://doi.org/10.1002/2015GL065271
  
\bibitem[{Mazelle et~al.}(2004)]{Mazelle04}
  Mazelle, C., D. Winterhalter, K. Sauer, J.G. Trotignon,
  M.H. Acuña, K. Baumgärtel, et al. (2004),
  Bow Shock and Upstream Phenomena at Mars,
  \textit{Space Science Reviews} 111, 115--181. https://doi.org/10.1023/B:SPAC.0000032717.98679.d0

\bibitem[{Ramstad et~al.}(2015)]{Ramstad15}
  Ramstad, R., S. Barabash, Y. Futaana, H. Nilsson, X.-D. Wang,
  and M. Holmstr\"{o}m (2015),
  The Martian atmospheric ion escape rate dependence on solar wind and
  solar EUV conditions: 1. Seven years of Mars Express observations,
  \textit{Journal of Geophysical Research: Planets},
  120, 1298--1309. https://doi.org/10.1002/2015JE004816

\bibitem[{Rojas-Castillo et~al.}(2019)]{Rojas18}
  Rojas-Castillo, D., H. Nilsson, and G. Stenberg Wieser (2019), 
  Mass Composition of the Escaping Flux at Mars: MEX Observations,
  \textit{Journal of Geophysical Research: Space Physics},
  Volume 123, Issue 10, pp. 8806--8822. https://doi.org/10.1029/2018JA025423

\bibitem[{Vignes et~al.}(2002)]{Vignes02}
  Vignes, D., M.H. Acuña, J.E.P. Connerney, D.H. Crider, H. Rème, and
  C. Mazelle (2002), Factors controlling the location of the Bow Shock at Mars, 
  \textit{Geophysical Research Letters},
  Vol. 29, No. 9, 1328. doi:10.1029/2001GL014513
\end{thebibliography}
\end{document}